\documentclass[12pt]{article}
\usepackage[fleqn]{amsmath}
\usepackage{amssymb}
\usepackage[dvips]{graphicx}
\usepackage{color}
\usepackage{tabularx}
\usepackage{algorithm}
\usepackage{algorithmic}
\usepackage{authblk}

\begin{document}
\title{Simulation of many-electron systems that exchange matter with the environment}
\author{Luigi Delle Site}
\affil{Institute for Mathematics, Freie Universit\"at Berlin, Germany}
\date{\vspace{-5ex}}
\maketitle
\begin{abstract}
The computational treatment of many-electron systems capable of exchanging {electrons and nuclei} with the environment represents one of the outermost frontiers in simulation methodology. The exchanging process occurs in a large variety of natural and artificially induced phenomena which are of major relevance to several leading fields of academic research and modern technology. In this progress report I will present an overview of problems in current materials science and chemical physics where the corresponding computational approaches require the concept of an electronic system with open boundaries. Quantum and Quantum/Classical computational techniques treat the exchange of electrons with the environment at different computational efficiency, conceptual rigorousness and numerical accuracy. The overall emerging picture shows a rich availability of interesting ideas, some with a higher weight on the pragmatic side, others with higher weight on the conceptual side; possible combinations, in perspective, may push the field much beyond its current frontiers.  
\end{abstract}

\section{Introduction}
Open systems that exchange matter with the environment represent a major challenge for theoreticians and simulators. In fact, the variation of the number of particles during the evolution of the system corresponds to a sudden change of the total amount of microscopic information that one must process and analyze within a consistent system-environment physical framework. In a previous work co-authored with Matej Praprotnik, \cite{physrep} an overview of theoretical principles and (mostly classical) simulation techniques for systems with open boundaries available in the literature has been discussed. In \cite{physrep} it was concluded {that when modeling matter classically, there is a satisfactory understanding of how to treat the case of varying the number of particles on a numerical level as well as on a conceptual level, with the corresponding dimensional change of the phase space. On the other hand, departing from a classical representation and instead modeling this process in the quantum case becomes far more complex.} In standard situations, e.g. in equilibrium, a classical particle arriving from the external environment into a system needs only to accommodate locally and does not modify abruptly the overall state of the system. The same concept does not hold in the case of quantum particles where the change in the number of particles completely redefines the quantum state of the system. Quantum particles are characterized by quantum correlations or better, by their entanglement, which implies that the most accurate knowledge of a system does not imply the most accurate knowledge of its parts \cite{kais,markusreiher,tecmer,ijqc}, thus the gain/loss of information for particles entering/leaving a subsystem (of specific interest) must be treated and interpreted with special care. In this work I will focus on the treatment of one specific, albeit relevant, class of quantum systems, that is many-electron systems that exchange particles, {i.e. electrons, or atoms/molecules (electrons + classical nuclei)}, with a large environment. The description of the environment and its coupling to the (sub)system are the two main ingredients for the construction of a computational procedure that can simulate the process of exchange. The environment can be considered with its full electronic structure, that is, if rigorously treated, the system of interest is merely a subsystem of a fully resolved quantum large system. In reality, pragmatic approximations are used to fully resolve the environment at a reasonable computational price. In general, the conceptual advantage of this model is that the electronic coupling (system-environment) is explicitly taken into consideration, and as a consequence the electronic correlations between the subsystem and the environment are, with some degree of accuracy, explicitly considered into the electronic properties of the subsystem. The computational disadvantage is that one needs to treat either large systems, which would be in most of the cases prohibitive for current computational resources, or a small environment which is likely to suffer from artifacts due to the reduced size.  Alternatively, the environment can be considered as an ideal statistical reservoir which is assumed to provide or adsorb particles according to the electronic chemical potential. It follows that the system is described as a quantum Grand Canonical ensemble without the need of explicitly calculating the electronic properties of the environment. The advantage in such a case is that one can focus only on the (sub-)system of interest, but the obvious disadvantage is that information regarding quantum correlations with the environment cannot be derived in any manner.
From the computational point of view, the two categories outlined above can be differentiated in terms of calculations at a fixed number of electrons (energy minimization of the system+environment) and calculations at a fixed chemical potential and a varying number of electrons for the system of interest.
The general conceptual framework outlined above leads us to the backbone of the paper. I have taken as guiding examples the theoretical and computational procedures employed in two relevant classes of applications: (i) in nanotechnology and (ii) in chemical and biological physics. More specifically, to the first category belong the subjects of nanoelectronics and electrochemistry, where the varying number of particles corresponds to the transport of electrons between  a (molecular) system and the environment acting as an electrode. To the second category belongs the field of solvation chemistry, where a molecular subsystem exchanges molecules with the large thermodynamic bath in which the subsystem is embedded, for example biomolecules in water. Here the exchange of electrons occurs through the exchange of molecules (i.e. electrons and nuclei) and {from a computational point of view, in order to make calculations more feasible, the environment shall be treated classically \cite{qmmmad2,qmmmad,cpcluigi}. In this case, the combination of quantum mechanics of electrons and the thermodynamics/statistical mechanics of the classical scale becomes particularly delicate \cite{physrep}.} Regarding the methods of calculation, among the electronic structure techniques, {Density Functional Theory (DFT) \cite{dft}, due to its intrinsically statistical mechanical structure, facilitates the extension of its principles to the treatments of electrons in various ensembles and at the same time it allows for the use of functionals beyond the total energy. The previous statement can be verified by consulting the seminal book of Parr and Yang \cite{bookdft} where DFT is derived in terms of density matrix and corresponding statistical ensembles}. As a consequence, DFT represents the most flexible and popular approach to the Grand Canonical treatment of electrons; a large amount of studies of chemical and physical systems where the DFT Grand Canonical approach has been used are present in the literature (see e.g. \cite{lozovoi,sprik,anatole,auer,arias,ayers1} and references therein). In general, all of the applications reported above use DFT, in various forms and in combination with other techniques, as the electronic structure method of preference. Beyond DFT, among the most advanced (wavefunction-based) electronic structure techniques, one finds high level quantum chemical methods in Fock space. In fact, this is a natural approach to create and destroy electrons in a system \cite{bochum}. In addition, Quantum Monte Carlo (QMC), given its stochastic/statistical nature  \cite{qmc-generic}, is flexible enough to treat the case of a varying number of particles and thus, treat the Grand Canonical ensemble \cite{qmc-size-a,qmc-size-b,qmc-size1,qmc-size2}. I will discuss the principles of the abovementioned QMC for one specific example, although it must be underlined that such an approach has not been explicitly used to treat the physical exchange of electrons, but has been employed mostly as a numerical trick that minimizes the size effects of the calculations{, i.e. it improves the convergence of calculated quantities to their corresponding value in the thermodynamic limit}. Nevertheless, as an electronic structure Grand Canonical technique, the QMC approach may be embedded in multiscale methodologies of the near future that will treat the physical process of varying the number of electrons and molecules. Finally, the paper is concluded with a discussion where the current ideas and related techniques are put in perspective. Possible combinations and modifications are suggested, which may be likely to optimize the methodology and hopefully push it beyond the current frontiers of applicability to physical systems.
\section{Electron flow: nanoelectronics and electrochemistry}
The passage of electrons from a molecule to its environment that acts as an electrode (and vice versa) is the prototypical situation occurring in nanoelectronics and electrochemistry. In nanoelectronics, the most popular case treated is that of a molecule that acts as a junction/bridge between two surfaces (metals or semiconductors) and allows the passage of a current of electrons from one surface to the other \cite{nanoel1,nanoelnic,nanoelrefpap,nanoel-advts}. In electrochemistry, a typical example is the flow of electrons between the reactants and a catalytic surface as a chemical reaction proceeds \cite{46ofarias}. In the next two sections, I will analyze how these two situations are treated at methodological level.
\subsection{Electron Transport in Molecular Junctions}
\begin{figure}[htbp]
\centering
\includegraphics[clip=true,trim=0.1cm 0cm 0cm 0.1cm,width=9cm]{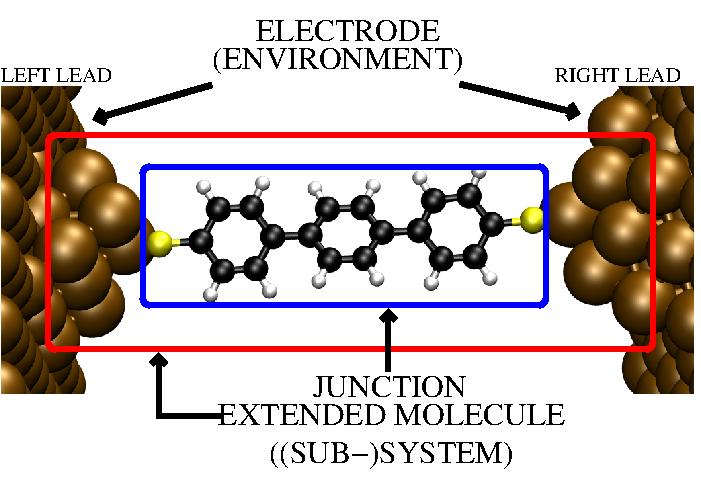}
\caption{Schematic illustration of a junction. Left and right lead are part of the environment (electrode), while the molecule is the system of interest through which the electrons flow. For numerical convenience, often the system of interest is the molecule and some of the metal atoms of the electrode, this approach goes under the name of ``extended molecule''.}
\label{junction}
\end{figure}

Figure \ref{junction} depicts the typical structure of a simple molecular junction attached to a right and left electrode.
{Differently from the other examples treated in this work (systems in equilibrium), the theoretical description of the transport of electrons from one electrode to the other is particularly challenging. In fact in such a case one deals with the more complex situation of non-equilibrium for a quantum many-body system.} In the analysis of the methodological aspects relevant to our current discussion, I will follow the recent perspective paper of Thoss and Evers \cite{nanoelrefpap}, which exhaustively traces the current state of the art in the field (see also references in the special issue about frontiers in molecular electronics \cite{spcnano}). They make a major distinction between {\it weakly correlated systems}, that is, systems where the energy transfer between charge carriers and the molecule can be neglected, and {\it strongly correlated systems}, where electron-electron or electron-phonon correlation have a significant impact. The weakly correlated regime, for its methodological aspects related to the system-environment treatment, is of particular interest to our specific analysis and will be discussed in the next section. For the strongly correlated regime I will not discuss the specific aspects of the strong correlations and its physical implications, instead I will restrict the attention to its treatment in the framework of a dynamic model of quantum open systems that exchange {electrons} with a reservoir.
\subsubsection{Weakly correlated regime}
The weakly correlated regime allows the description of a molecular junction within the approximation of quasi-static scattering potential and is valid in the regime of linear response, that is, the transport is treated as an elastic process. Within this picture, the electronic properties of interest can be calculated by Kohn-Sham Density Functional theory (KS DFT)\cite{dft}. The current-voltage characteristic, following the Landauer formula is \cite{nanoelnic,landauer,ratner}:
\begin{equation}
  I(V)=\frac{2e}{h}\int_{-\infty}^{\infty}T(E)[f_{L}(E)-f_{R}(E)]dE
\end{equation}
where $V$ is the applied voltage and $E$ the energy levels, $f_{L,R}(E)=f(E-\mu_{L,R})$ is the Fermi distribution function of the lead at the corresponding chemical potentials $\mu_{L,R}$. In this context, the key quantity is the (molecular) transmission function, $T(E)$ which can be written as (see also \cite{koentopp}):
\begin{equation}
T=tr {\bf \Gamma_{L}\mathcal{G}_{V}\Gamma_{R}
  \mathcal{G}^{\dagger}_{V}}.
\end{equation}
The quantities involved in the formula are: ${\bf \mathcal{G}}^{-1}_{V}={\bf \mathcal{G}}^{-1}_{0V}-{\bf \Sigma}_{L}-{\bf \Sigma}_{R}$, where the resolvent matrix is:
\begin{equation}
  {\bf \mathcal{G}}^{-1}_{0V}({\bf x},{\bf x}^{'},E)=\sum_{n}\frac{\phi_{n}({\bf x})\phi^{*}_{n}({\bf x}^{'})}{E-\epsilon_{n}+i\eta/2}
\end{equation}
the sum is over the KS energies and orbitals, $\epsilon_{n}$ and $\phi_{n}$ for the uncoupled molecule calculated with a standard DFT approach. ${\bf \Sigma}_{L,R}$ are the self-energies, so that in the approximation of non-interacting particles, ${\bf \Gamma_{L,R}}=i({\bf \Sigma}_{L,R}-{\bf \Sigma}^{\dagger}_{L,R})$. The self-energy is the coupling term between molecule and leads, and can be formulated in terms of a hopping matrix: ${\bf \Sigma}_{L,R}={\bf t}_{L,R}{\bf g}_{L,R}{\bf t}^{\dagger}_{L,R}$. The elements of the hopping matrix are defined as: $t_{L,R}({\bf X}N, {\bf x}n)$, describing the hopping process of an electron in an orbital $N$ of an atom at position ${\bf X}$ of the molecule to an orbital $n$ of an atom at position {\bf x}, in the lead. These elements are approximated by their bulk values and are calculated using independent DFT simulations of the lead (i.e. usually a large metal cluster without the molecule) and the molecule. Finally, the surface Green's function, $<{\bf x}n|{\bf g}_{L,R}|{\bf x}^{'}n^{'}>$, that is the effect of broken translational symmetry  on the density of states of the metal occurring at the interface plane perpendicular to the lead-molecule direction, is also calculated in {the same DFT simulation of the metal cluster used to find the elements of the hopping matrix.} This description reports the very essential aspects to give an idea of the methodological procedure used in the field; there are of course technical improvements (e.g. ``the extended molecule'' model, that is the molecule and some atoms/layers of the leads) and further conceptual developments of this basic picture. Advantages, limitations and open problems are well described and discussed in Reference \cite{nanoelrefpap}. For the current focus the relevant methodological aspects are: (i)  the model of environment/lead, is fully resolved in its electronic structure, although in the pragmatic approximation of large clusters, (ii) its coupling to the system/molecule is done through the hopping process by which electrons are exchanged. This latter aspect requires the construction of a matrix obtained by (iii) modeling  the system of interest/junction as an ``isolated molecule'' treated, as for the bulk of the metal, with standard DFT approaches at a constant number of electrons. The approach underlined in this section makes use of a static picture where the dynamics of electron exchange between molecule and lead is expressed through the probabilistic event of the hopping. An interesting methodological variation for the calculation of $T(E)$ where the leads are characterized only by their chemical potentials (given as an input), without the need of knowing their electronic structure, has been proposed by Arnold, Weigend and Evers \cite{quasi-gc}. They develop a general scheme, which despite requiring standard quantum chemistry calculations for the bridging molecule (or extended molecule), in practice gives as a result electronic properties calculated at fixed chemical potentials (of the left and right lead) and a variable number of electrons. This is an effective and computationally robust way to realize the Grand Canonical Ensemble in an implicit manner from a series of (properly looped) standard fixed-particle quantum calculations. Its potential application can go far beyond the case of electron transport and can certainly be used for systems in equilibrium that will be described later on. Beyond the static approaches discussed so far, there exists a class of methods that explicitly considers the electron dynamics and can treat also the {\it strongly correlated} regime. They are based on the Liouville equation for the density matrix of the subsystem of interest (reduced density matrix of the junction). In essence, such an equation describes the dynamical behaviour of electrons in a molecule considered as an open quantum system; such an approach is described in the next section.
\subsubsection{Electron Dynamics: Liouville Equation for the Density Matrix and related methods}
In an interesting work of Emch and Sewell \cite{emsew}, the Liouville equation for the time evolution of a quantum subsystem (S) embedded in an ideal bath (R) is rigorously derived in terms of the evolution of the density matrix of the subsystem. The interesting aspects for the focus of this work is that the authors suggest a coupling term between the S and R also for the case in which the bath is a reservoir of particles (in addition to energy). The essence of the Emch-Sewell model is the filtering of the microscopic degrees of freedom of R through the Zwanzig projector \cite{zwanzig}. The process of filtering leads to effective actions of such microscopic degrees of freedom in terms of actions of statistically averaged macroscopic quantities.
The standard Liouville-Neumann equation: $\frac{d}{dt}\rho=-iL\rho$, with $\rho$ the density matrix of the system and $L=[H,*]$ the Liouville operator with Hamiltonian, $H$, is transformed through the projector operator $\mathcal{P}$ into the equivalent master equation for $\mathcal{P}\rho$:
\begin{equation}
\frac{d}{dt}\mathcal{P}\rho(t)+i\mathcal{P}L\mathcal{P}\rho(t)+\int_{0}^{t}dt^{'}\mathcal{P}L(I-\mathcal{P})\mathcal{U}(t-t^{'})(I-\mathcal{P})L\mathcal{P}\rho(t^{'})=0
\label{eq1}
\end{equation}
$I$ is the identity operator and $\mathcal{U}(t)=exp[-i(I-\mathcal{P})L(I-\mathcal{P})t]$. S and R are initially uncorrelated/independent $\rho(0)=\rho_{R}(0)\otimes \rho_{S}(0)$. The initial state of R is given by the measurement of the set of macroscopic variables, thus $\mathcal{P}\rho(0)=\rho(0)$ and since the average/macroscopic quantities of the reservoir are time independent, then $\mathcal{P}\rho(t)=\rho_{R}(0)\otimes \rho_{S}(t)$ thus Eq.\ref{eq1} can now be written as a self-contained equation for S:
\begin{equation}
\rho_{R}(0)\otimes\left(\frac{d}{dt}+i L_{eff}^{S}\right)\rho_{S}(t)=-\int_{0}^{t}dt^{'}\mathcal{K}(t-t^{'})\rho_{R}(0)\otimes\rho_{S}(t^{'})
\label{eq2}
\end{equation}
considering the trace with respect to R on both sides one obtains the master equation for $\rho_{S}(t)$:
\begin{equation}
\left(\frac{d}{dt}+i L_{eff}^{S}\right)\rho_{S}(t)=-\int_{0}^{t}dt^{'}\mathcal{K}^{S}(t-t^{'})\rho_{S}(t^{'}).
\label{eq3}
\end{equation}
$\mathcal{K}^{S}(t)\rho_{S}(t)=Tr_{R}\{\mathcal{K}(t)\rho_{R}(0)\otimes \rho_{S}(t)\}$. As anticipated above, the interesting point for our focus is that $ L^{S}_{eff}$ and $\mathcal{K}(t)$  can be used to define the exchange of particles between R and S. The Hamiltonian of the system can be divided in three parts: $H=H_{R}+H_{S}+H_{I}$, the relevant part is the latter term, that is the Hamiltonian of interaction between R and S. Correspondingly, one has  the related Liouville operators $L_{R}, L_{S}, L_{I}$. 
The authors define $H_{I}=\int_{\Omega_{R}}dx\int_{\Omega_{S}}dy V(x,y) J_{R}(x)\otimes J_{S}(y)$, with $x,y$ configuration coordinates of R and S respectively, $\Omega_{R}, \Omega_{S}$ the volumes occupied and $J_{R}(x),J_{S}(y)$ operators acting on the Hilbert subspace of R and S respectively. Such operators represents intensive variables, for example particle number, which could be function of the creation and annihilation operators for particles in R and in S. The intensity of their action is regulated by $V(x,y)$, a potential of (direct) interaction between R and S (e.g Voltage in a junction-lead system).
Furthermore, $L^{S}_{eff}=L_{S}+L^{S}_{I}$, with $L^{S}_{I}\rho=[V_{S},\rho]$ and $V_{S}=\int_{\Omega_{S}}dy\langle V(y)\rangle_{0} J_{S}(y)$. The kernel is defined as follows:
\begin{equation}
  \mathcal{K}(t)=\mathcal{P}\mathcal{U}_{S}(t)L_{I}(t)(I-P)\mathcal{U}^{'}(t) L_{I}\mathcal{P}
\end{equation}
with
\begin{equation}
  \mathcal{U}_{S}(t)=exp\{-i L_{S}t\}$, $L_{I}(t)=exp[i(L_{R}+L_{S})t] L_{I} exp[-i(L_{R}+L_{S})t]
\end{equation}
and
\begin{equation}
  \mathcal{U}^{'}(t)=exp\{-\int_{0}^{t}dt^{'} (I-\mathcal{P}L_{I}(t^{'})(I-\mathcal{P})\}.
\end{equation}
{The treatment is (in principle) exact, but the memory kernel , $\mathcal{K}_{S}(t-t^{'})$, cannot be analytically determined thus it must be approximated. As suggested in \cite{nanoelrefpap}, recent research has brought substantial advances in the field \cite{memory1,memory2}.}
A similar formulation in which the Liouville-Neumann equation considers a stochastic coupling to a bath is the so-called Kossakowski-Lindblad equation \cite{linda,kosslinda}:
\begin{equation}
\dot{\rho(t)}=L(\rho)=i[H,\rho]+\frac{1}{2}\sum_{j}([\mathbb{L}_{j}\rho,\mathbb{L}^{+}_{j}]+[\mathbb{L}_{j},\mathbb{L}^{+}_{j}\rho])
\label{lineq}
\end{equation}
with $H$ the system Hamiltonian, $\mathbb{L}_{j}$,$\mathbb{L}_{j}^{+}$ operators that carry the interaction of the system with a reservoir (Lindblad operators). The term $\sum_{j}([\mathbb{L}_{j}\rho,\mathbb{L}^{+}_{j}]+[\mathbb{L}_{j},\mathbb{L}^{+}_{j}\rho])$, gives Eq.\ref{lineq} the form of a rate equation (quantum jumps of the system under the action of the environment), $[\mathbb{L}_{j}\rho,\mathbb{L}^{+}_{j}]$ and $[\mathbb{L}_{j},\mathbb{L}^{+}_{j}\rho]$ can be interpreted as transition rates between two events (e.g. particles exchange with a reservoir and transition from $N$ to $N^{'}$). The use of the creation and annihilation operators inevitably leads to the treatment of the problem in Fock space, within standard one-particle orbitals of quantum chemistry, or Bloch states and plane waves \cite{kosov,neuhauser}, further routes to describe the electron transport originating from the master equation and references for specific determinations of the memory kernel are discussed in Ref.\cite{nanoelrefpap,tamar}.
It must be reported that methods based on the direct calculation of the time evolution of the wavefunction of the system, rather than its density matrix, are also present in the literature, however in such a case, the wavefunction must be calculated for the whole system (molecule+leads). In order for this calculation to be computationally feasible, the leads must be approximated by a finite system \cite{thosswf}. The most popular approach in such a context is the time-dependent Hartree method (MCTDH) \cite{lubic,vanleuven}. For electron transport, the exchange of electrons between system (molecule) and environment (leads) occurs through the change of orbital occupations (from lead to molecule and molecule to lead) \cite{thosswf2}. In such case correlations are explicitly taken into account, however, the computational effort may be sizable \cite{nanoelrefpap}. Beyond electron transport, within the same wavefunction approach, attempts to restrict the treatment only to a subsystem and add the rest (environment) as e.g. a sink of electrons (i.e. for treating ionization processes), led to different technical and conceptual problems; the most relevant is the fact that the modified Hamiltonian, with a sink: $H+i\Gamma$ with $\Gamma$ a one body potential vanishing on the domain of the molecule, is a non-Hermitian operator \cite{kvaal}. Overcoming such a limitation implies the fulfillment of the requirement that the quantum dynamical semigroup for the time evolution must be trace preserving, Markovian and strictly positive. In practice, one follows the application of the Gorini-Lindblad theorem \cite{gor,linda}, and thus arrives at the  Kossakowski-Lindblad equation discussed before. The previous statement implies that the density matrix, when the number of electrons is varying in time, is the proper quantity to consider rather than the direct wavefunction of the system. The use of the Liouville equation discussed so far implies two distinct computational strategies for the partitioning and definition of system and environment. In the approach of the projector operator, the density matrix should, in principle, be defined for the whole space and the physics of the subsystem is obtained by the action of Zwanzig projector operator. It is clear that while the accuracy of the results would be very high, its computational costs are likely to be prohibitive for systems of reasonable size (see also comments about Fock space), and of course reasonable approximations are possible and needed (see e.g. \cite{nanoelrefpap} and references therein). The Liouville equation in the approach of Kossakowski-Lindblad, requires instead only the definition of a transition probability without the necessity of explicit calculations of the electronic properties of the reservoir. Though, of course, its accuracy is directly linked to the construction of a physically well funded stochastic term, for example from transition rates determined experimentally. 
In general, as underlined before, the treatment of a problem in Fock space is the natural approach to describe the passage of electrons from one system to another, but in practice the computation may be difficult. In fact, it requires the construction of the density matrix through the treatment of a large number of states to which {electrons} can be located according to the action of the creation and annihilation operators. This means that one needs a predefined reasonable ``active space'' \cite{bochum}. If the explicit time dependence of the electron flow is not of primary relevance, an alternative to the use of Fock space is given by DFT treated in the Grand Canonical ensemble. Calculations for electrochemistry are often done following this approach, and this subject will be treated in the next section.
\subsection{Grand Canonical Density Functional Theory and its application to Electrochemistry}
\begin{figure}[htbp]
\centering
\includegraphics[clip=true,trim=0.1cm 0cm 0cm 0.1cm,width=10cm]{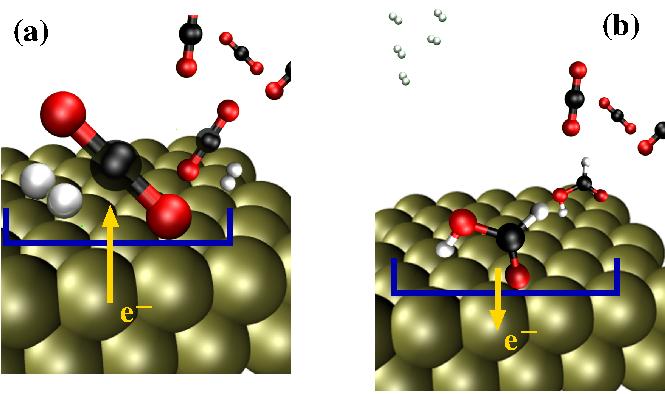}
\caption{Schematic illustration of an electrochemical process at a metal surface (reduction of formic acid). Panel (a), $H_{2}$ and $CO_{2}$ molecules solvated in water react, through the supply of electrons from the surface, and produce formic acid. Next electrons are released and adsorbed by the surface, Panel (b). In this case the system of interest are the  reactants, $H_{2}$ and $CO_{2}$ molecules and the product, the formic acid molecules, while the environment consists of the solvating bath of water and the metal surface  which acts as a reservoir of electrons.}
\label{electro}
\end{figure}

Electrochemical systems can typically be schematized as an electrolyte, usually consisting of ions in a solvent, and a surface, acting as an electrode, which promotes an electrochemical reaction \cite{46ofarias,cat1}. Electrons are absorbed from or injected into the solution as the reaction proceeds (see also Figure \ref{electro} for a pictorial example).
From the point of view of simulation, modeling such systems are extremely challenging: the full quantum treatment of the system is computationally prohibitive, while a pragmatic partitioning of the total system as subsystem of interest (where the chemical reaction occurs), and the environment (electrolyte and electrode), allows multiscale simulation techniques to be applied with success \cite{arias}. Later on I will describe a specific model/example that treats the electrolyte, but for the moment, the interesting aspect is the treatment of the electrode as a Grand Canonical reservoir of electrons for the (sub)system of interest. In fact, such a view can be generalized to the problem of a charged surface or a surface in an external electric field which can be treated as a system at constant chemical potential, $\mu$ and varying number of electrons $N$ (for methodologically-oriented work see e.g. \cite{ali,thomas,auer,arias,ayers1}). The use of DFT calculations with constant $\mu$ and varying $N$  leads to the introduction of the (exact) Helmholtz free energy, $A$, for interacting electrons at finite temperature in an external potential  $V({\bf r})$. $A$ satisfies the so-called Hohenberg-Kohn-Mermin variational theorem \cite{dft,mermin} (see also Reference \cite{arias}, of which I will follow the formalism):
\begin{equation}
A=\min_{\{n({\bf r})\}}\left(A_{HKM}[n({\bf r})]+\int V({\bf r}) n({\bf r}) d{\bf r}\right)
  \label{A1}
\end{equation}
with $A_{HKM}[n]$ the universal free energy functional of the electron density $n({\bf r})$, for any external potential $V({\bf r})$ (in atomic units).
As for the universal energy functional of Hohenberg and Kohn, $A_{HKM}[n({\bf r})]$ is not known and can be only approximated:
\begin{equation}
  A_{HKM}[n]=A_{ni}[n]+E_{H}[n({\bf r})]+E_{XC}[n({\bf r})]
  \label{A2}
\end{equation}
where $A_{ni}[n]$ is the non-interacting free energy, $E_{H}[n]$ is the Hartree term and $E_{XC}[n]$ is the exchange and correlation energy. The relevant quantity here (compared to standard DFT) is:
\begin{equation}
  A_{ni}[n]=\min_{\{(\psi_{i}({\bf r}),f_{i})\to n({\bf r})\}}\sum_{i}\left(\frac{f_{i}}{2}\int |\nabla\psi_{i}({\bf r})|^{2}d{\bf r}- T S(f_{i})\right)
  \label{A3}
\end{equation}
that is, the minimum over all the single-particle orbitals $\psi_{i}({\bf r})$ and the corresponding occupation factors $f_{i}\in [0,1]$ which leads to the electron density: $n({\bf r})=\sum_{i}f_i|\psi_{i}({\bf r})|^{2}$. $T$ is the temperature and $S(f_{i})$ is the single-particle entropy:
\begin{equation}
  S(f)=-f\ln f-(1-f)\ln(1-f).
\end{equation}
The minimization leads to the single-particle KS equations:
\begin{equation}
    -\frac{\nabla^{2}}{2}\psi_{i}({\bf r})+V_{KS}({\bf r})\psi_{i}({\bf r})=\epsilon_{i}\psi_{i}({\bf r})
\end{equation}
for the stationarity w.r.t. $\psi_{i}({\bf r})$ and to the Fermi Occupation condition:
\begin{equation}
  f_{i}=\frac{1}{1+e^{\frac{\epsilon_{i}-\mu}{T}}}
\end{equation}
for the stationarity w.r.t. $f_{i}$, with $V_{KS}({\bf r})=V({\bf r})+\frac{\delta}{\delta n({\bf r})}(E_{H}[n]+E_{XC}[n])$. For standard DFT, $\mu$ is a Lagrange multiplier and the fixed number of {electrons} is assured by the condition: $\sum_{i}f_{i}=N$. In a Grand Canonical ensemble instead, the free energy to minimize is not $A$ but the Grand Free Energy:
\begin{equation}
  \Phi=A-\mu N
\end{equation}
In such a case, for the optimization procedure applied above, one modification is required, that is the Lagrange multiplier term $-\mu(\sum_{i}f_{i}-N)$ is replaced by $-\mu\sum_{i}f_{i}$. Such modification removes the constraint on fixed $N$ and implements the Legendre transformation from $A$ to $\Phi$; $\mu$ is now given as an input while $N$ is variable.
The grand potential of the electrons can be expressed also in a different formalism, for example following Alavi {\it et al.} \cite{ali}:
\begin{equation}
  \Phi[n({\bf r})]=-\frac{2}{\beta}\ln det\left(1+e^{-\beta(\mathcal{H}-\mu)}\right)-\int d{\bf r} n({\bf r})\left(\frac{\phi({\bf r})}{2}+\frac{\delta\Phi_{xc}}{\delta n({\bf r})}\right)+\Phi_{xc}
\end{equation}
with $\Phi_{xc}$ the finite-temperature exchange-correlation grand potential, $\phi({\bf r})$ the Hartree potential, $\beta=\frac{1}{k_{B}T_{e}}$ the electronic temperature parameter, $\mathcal{H}=-\frac{1}{2}\nabla^{2}+V({\bf r})$ the one-electron Hamiltonian, and the effective density-dependent potential, $V({\bf r})=V_{ext}({\bf r})+\phi({\bf r})+\frac{\delta\Phi_{xc}}{\delta n({\bf r})}$, with $V_{ext}({\bf r})$ the external potential. The exponential form can then be efficiently evaluated through the Trotter approximation as a product of $P$ high temperature matrices: $e^{-\beta\mathcal{H}}=\left(e^{-\beta\mathcal{H}/P}\right)^{P}$, with $P$ a large integer, so that $\epsilon=\beta/P$ is small and one can write: $e^{-\epsilon(K+V)}=e^{-\epsilon V/2}e^{-\epsilon K}e^{-\epsilon V/2}+\mathcal{O}(\epsilon^{3})$; based on such an idea, it was possible to devise efficient linear system-size scaling schemes \cite{parr1,parr2,thomas}.
In general, compared to fixed $N$ calculations, those at constant $\mu$ carry several technical problems. For example, the large fluctuations of $N$ (and $n({\bf r})$) at the initial stage, the need of compensating charges because  of the finiteness of the slabs representing the surface and periodic boundary conditions \cite{lozovoi,auer,arias}, or the fact that $E_{XC}[n]$ are defined for integer numbers of electrons \cite{perdew,vuilleumier,weitao}. However, technical solutions were made available and thus the method can be routinely used nowadays for applications. The approach outlined above makes possible the introduction of a reservoir of electrons that adds or removes {such} particles as, e.g., a chemical reaction on a surface/electrode proceeds, so that our system of interest consists of the reactants while the surface represents the (active) environment. However, as underlined before, the effect of the solvent often plays a key role and its quantum treatment would be computationally prohibitive. The most popular approximation consists of employing continuum solvation models \cite{cont1,cont2}, however an interesting proposal based on the so-called joint density-functional theory (JDFT) \cite{jdft1,jdft2} has been elaborated in the context of Grand Canonical DFT reported above \cite{arias}. JDFT consists of a variational theorem similar to the Hohenberg-Kohn theorem which allows for the description of the free energy of a solvated system in terms of the electron density $n({\bf r})$ of the solute and of nuclear densities for the solvent, $N_{\alpha}({\bf r})$, where $\alpha$ indicates a nuclear species:
\begin{equation}
  A=\min_{\{n({\bf r}), N_{\alpha}({\bf r})\}}\left(A_{JDFT}[n({\bf r}),N_{\alpha}({\bf r})]+\int V({\bf r})n({\bf r})d{\bf r}+\sum_{\alpha}V_{\alpha}({\bf r})N_{\alpha}({\bf r}) d{\bf r}\right)
  \label{jdft1}
\end{equation}
similarly to the Hohenberg-Kohn-Mermin functional, $V({\bf r})$ is the external electron potential, $V_{\alpha}({\bf r})$ is the external potential for the nuclei of the solvent and $A_{JDFT}[n({\bf r}),N_{\alpha}({\bf r})]$ is a universal functional independent of the two potentials and can be separated as:
\begin{equation}
  A_{JDFT}[n,N_{\alpha}]=A_{HKM}[n]+A_{diel}[n,N_{\alpha}].
\end{equation}
The solvent is described in terms of average density and not in terms of individual atomistic configurations which would imply expensive sampling computations with e.g. molecular dynamics or Monte Carlo, while the system of interest is treated at full quantum (Grand Canonical) accuracy. However, often the solvent and the corresponding configuration space that it can access are of crucial importance for the chemical or physical event of interest and thus the continuum approximation is far too drastic and it requires an explicit molecular treatment of the solvent \cite{bulo-matter}. In general, the region of interest is usually localized in space, e.g. the solvation region of a solute, thus the quantum accuracy may be needed only in a small portion of the system while the rest of the system can be treated at a coarser (classical) level. This is a typical scenario for, e.g. biophysical systems such as proteins and membranes in water, whose relevance for the current academic research as well as for technology has stimulated the development of multiscale computational methodologies, such as Adaptive Quantum Mechanics/Molecular Mechanics (A-QM/MM) \cite{qmmmad} and Adaptive Resolutions Simulation (AdResS) \cite{physrep}. In such cases, the environment is a classical system and provides/takes molecules from the quantum region, thus the exchange of electrons is strictly related to the exchange of entire molecules. This kind of approach is the subject of the next section. 
\section{Molecules in Solution: Exchange of Molecules between system and reservoir}
Many interesting systems in chemical physics and in particular in biochemistry are characterized by events that happen locally, e.g. in solution. In such a context, the interesting event takes place in a specific subregion of the system (where electrons play a key role) embedded in a larger thermodynamic bath of, e.g., solvating molecules. The proper (full quantum/electronic treatment) in space and time, of such a class of systems  is in most of the cases prohibitive; however, a decisive step forward was taken following the idea of space partitioning of Warshel and Karplus \cite{warschkarplus} and of Birge, Sullivan and Kohler \cite{birge},  that is, each region is treated at the most convenient resolution, i.e. quantum if electrons play a major role or coarser otherwise. Later on, the idea of partitioning was extended further by Warshel and Levitt \cite{warshlev}, who described the enzymatic reaction of lysozyme; their seminal work is nowadays recognized as the first example of the so-called Quantum Mechanics/Molecular Mechanics (QM/MM) method. This method has represented a technical revolution in the field because it allows for the treatment of systems that, up to that moment, were thought to be intractable at quantum level. For this reason, the pioneers of the technique, Martin Karplus, Michael Levitt and Arieh Warschel were awarded the Nobel Prize in Chemistry in 2013. Today, QM/MM represents a very precious computational tool for tackling problems in various field of chemistry and physics with applications in a broad range of subjects and disciplines (see e.g. \cite{roth,thiel}). However, in the standard QM/MM method, the quantum region and the classical region are fixed, that is, there is exchange of energy, but not of matter. The physical implication is that the QM region represents an artificial Canonical ensemble whose accuracy increases by increasing the size of the QM region and checking that the effect of the environment is negligible. The most recent development in the field attempts to technically remove the constraints of a fixed number of molecules in the QM region and thus, allow the exchange of molecules between the quantum and the classical region (see e.g. \cite{qmmmad}).
I have already mentioned such an approach, referring to it as: ``Adaptive Quantum Mechanics/Molecular Mechanics (A-QM/MM)''; its advantages and limitations will be discussed in the next sections. In parallel, within the context of classical multiscale simulation, (molecular) adaptive  resolution methods for classical systems have been developed \cite{physrep}. Typically such methods concurrently couple regions at (classical) atomistic resolution with regions at coarse-grained resolution in such a way that when a molecule passes from one region to the other, it slowly changes resolution without creating sizable physical artifacts. 
The next frontier for classical adaptive resolution methods is the extension to molecules with electrons. In this context, a computational framework that I have recently proposed, electronic Quantum Adaptive Resolution Simulation (el-QM-AdResS) \cite{cpcluigi}, where the classical adaptive resolution technique is combined with the electronic scale, is discussed in the following sections. Finally, a discussion of the similarities, differences and possible synergies of A-QM/MM and el-QM-AdResS concludes the chapter.
\subsection{A-QM/MM: Partitioning the system on-the-fly}
\begin{figure}
   \centering
   \includegraphics[width=0.75\textwidth]{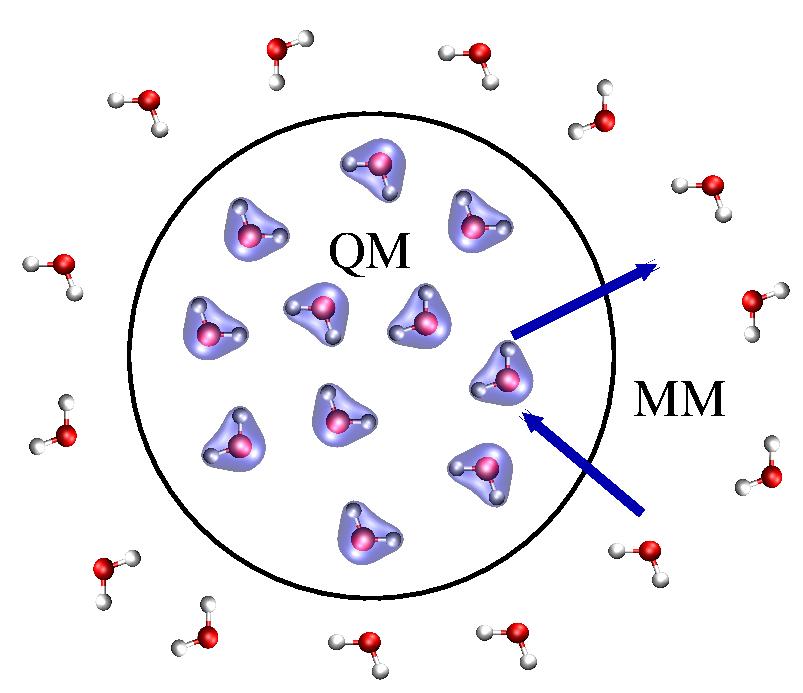}
   \caption{Schematic representation of the A-QM/MM method. Molecules that at time $t$ are in the QM region are treated at quantum mechanical level, molecules outside, in the MM region, are treated with standard classical force-field methods. {The example shown here involves water molecules: the quantum molecules are schematized as nuclei around which the electron density is distributed and the classical molecules are schematized as standard atoms using a classical MM force field.}}
 \label{abrupt-a-qmmm}
 \end{figure}

\begin{figure}
   \centering
   \includegraphics[width=0.75\textwidth]{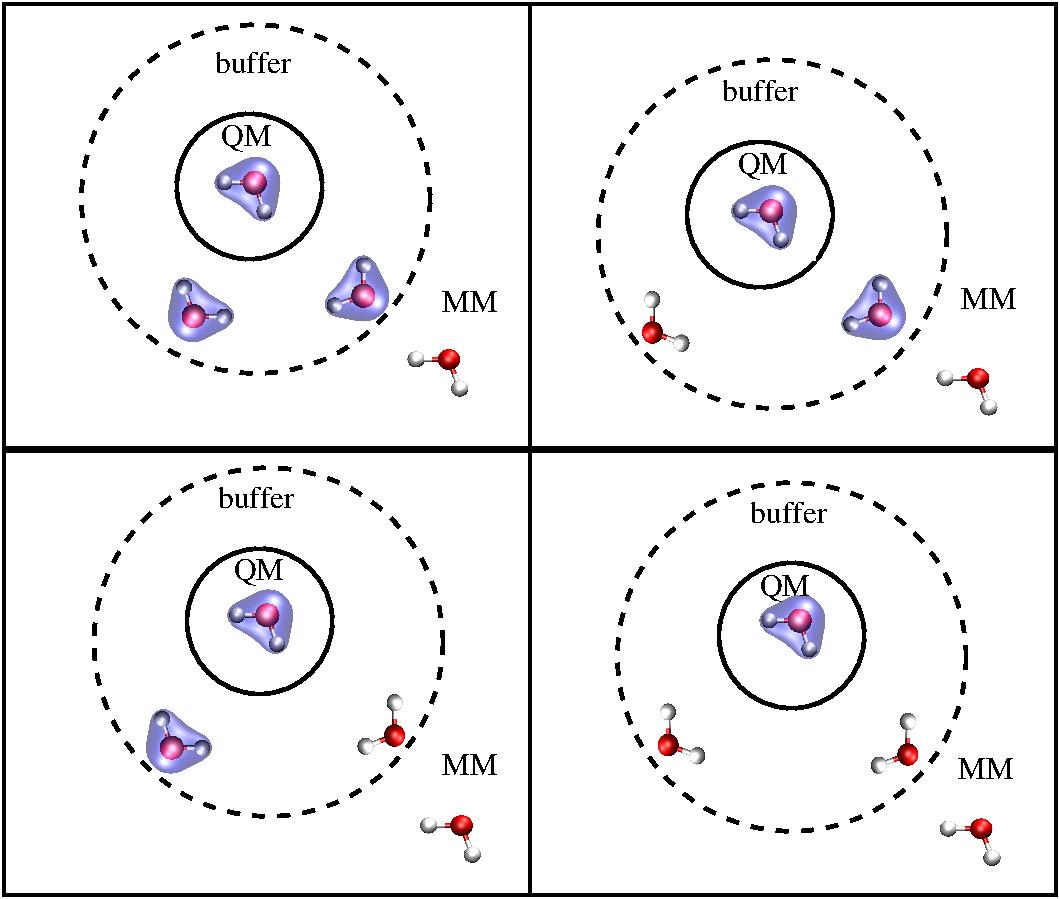}
   \caption{Example of the role of the buffer and the corresponding partitioning schemes for $M=4$, for a system of water molecules. As in the previous figure the quantum mechanical character is indicated by the electron density distributed around the nuclei.}
 \label{M-part}
 \end{figure}

In the following discussion, I will take as a main reference the exhaustive review by Hai Lin and collaborators \cite{qmmmad} where the state of the art of the field is critically analyzed (for an additional complementary view see also Ref.\cite{qmmmad2}). A-QM/MM, as anticipated above, is an extension of the standard QM/MM method. The essence of A-QM/MM consists of an on-the-fly reclassification of QM and MM atoms/molecules during the simulation so that the molecules included in the QM or in the MM region can be dynamically updated according to the evolution of the system in space. A straightforward approach consists of an ``abrupt'' scheme of reclassification (see Figure \ref{abrupt-a-qmmm}): classical molecules that at time $t_{0}$ lie at the border of the QM region may directly enter into the QM region at time $t_{1}=t_{0}+\Delta t$, ($\Delta t$ time step of the simulation) and be reclassified as QM molecules for the calculations of the next time step; equivalently, QM molecules lying at the border with the MM region and moving at time $t_{1}$ into the MM region are reclassified as classical for the calculations of the next time step.
It was found out that such a sudden change implies the hopping between different energy surfaces whose discontinuity causes numerical instabilities and artificial results \cite{heydenqmmm}. For such a reason, the latest generation of A-QM/MM methods are centered around more involved computational schemes based on the introduction of a buffer region. At each time-step of the simulation,
 the buffer region is partitioned in different subsets and the standard
 QM/MM interactions are defined for all possible subsets (see Figure \ref{M-part}). Next, for each partition, the molecules of the corresponding subset of the buffer are included in the QM region.
 Finally, for each of these ``extended'' QM regions a
 standard QM/MM calculation is done. {The total potential is then defined as a
 weighted average of the individual potentials from each individual simulation}:
\begin{equation}
  U({\bf r})=\sum_{i}^{M}f_{i}({\bf r})U_{i}({\bf r}).
  \label{qmmmint}
\end{equation}
Here, each $U_{i}({\bf r})$ corresponds to one of the $M$ partitions of the system in a group of QM
 molecules and a group of MM molecules and $f_{i}({\bf r})$ is a switching function depending on the coordinates of the single molecules. The switching function is constructed with the intuitive but well justified idea  that the quantum energy of molecules in the buffer at larger distances from the center of the region of interest (active site) contributes less than the energy of molecules which are closer to the active site. This is the essential technical and conceptual point of the method. In fact, the dynamical evolution produced by the energy surface, can now be updated and it
 creates a new configuration on which, in turn, the partitioning step and the corresponding weighted potential (or forces) are applied for the time evolution of the next time step.  From the technical point of view, the empirical approach to the smoothing of the coupling at the interface (either via potential or via inter-molecular forces) turned out to be very powerful and, in my view, can be considered a very relevant step forward in the development of truly multiscale simulation technologies for condensed matter and chemical physics, as testified by a large number of successful applications \cite{bulos1,nielsen,csanys,watanabe,lins1,lins2} (see also additional references in the topical reviews \cite{qmmmad2,qmmmad}). Yet, Lin and coworkers point out a series of technical problems, but above all conclude that despite the large number of encouraging results, A-QM/MM cannot be considered a truly predictive tool since its results always require a case-by-case validation, i.e. reproduction of some reference results (from experiment or larger QM/MM calculations). {In the next section, I will discuss conceptual aspects that are mandatory for the construction of a A-QM/MM approach with automatic {\it a priori} physical control criteria that can transform A-QM/MM into a predictive tool without the need for external validation.}
 \subsection{A-QM/MM: Physical validation and the necessity of a Grand Canonical view}
 In Reference \cite{qmmmad}, it is explicitly stated that, in principle, the QM region can be made as small as possible. This statement is certainly true from a technical point of view, however it carries the first conceptual bug of the current QM/MM methods (in general). In fact, if one is interested in a realistic quantum description of a specific region of interest, then the coupling energy between the QM region and the MM region must be negligible compared to the energy of the QM region, otherwise the electronic spectrum is essentially determined by the classical part. It is obvious that in QM/MM the electronic wavefunction and corresponding energy spectrum of the QM region cannot be the same as if the region was in a full QM environment. In good approximation, this situation holds only in the case of localized systems/properties, that is, when there is a negligible coupling energy with the environment (i.e. a reasonably separable Hamiltonian). This would be the first internal criterion of control for the validity of a QM/MM calculation. If we now take a step forward and move to the A-QM/MM approach, there are further conceptual problems. A very relevant one concerns the physical validity of the approach used in the partitioning scheme. Specifically: as said before, at quantum mechanical level, in terms of wavefunctions and electronic spectra, an open QM region embedded in a MM reservoir, in general, cannot be equivalent to a QM region embedded in a full QM reservoir. The equivalence can only hold from the quantum statistical mechanics point of view, if (and only if) the QM region of the A-QM/MM, is interpreted as a Grand Canonical region at the same macroscopic (thermodynamic) and electronic chemical potential of a QM region embedded in a full quantum environment.
 {In this respect, the concept of {\it ``Nearsightedness of electronic Matter''} introduced by W.Kohn \cite{nearkohn, nearkohn2}, provides a clear physical principle for the definition of a subsystem embedded in a larger environment. In fact it states that, at fixed chemical potential, local electronic properties depends on the effective external potential (i.e. the environment) only at nearby points. The effects of the external potential beyond a certain distance are negligible for the local properties. Furthermore, in a recent work of Fias, Heidar-Zadeh, Geerlings, and Ayers, it has been shown that the response kernel to the environment is local at constant chemical potential, thus in any partitioning scheme one should use the concept of constant chemical potential \cite{ayers-gerl}.}
 To my knowledge, the current partitioning schemes of A-QM/MM do not use {any of the criteria of control listed above}.
Moreover, in a recent work  by Miranda-Quintana and Ayers about electronic systems with a varying number of {electrons} in DFT, it is  discussed the possibility that interpolations of property-values between electron numbers is not consistent with a physical ensemble average \cite{quintana-ayers}.\\
 Their finding may be extended to the current discussion and would imply that any average property obtained by averaging over the different $M$ partitions of the A-QM/MM system, where each calculation is done minimizing the energy $E_{i}$ at a given (fixed) number of electrons $N_{i}$, is inconsistent with statistical mechanics.\\ 
 An extension of the A-QM/MM method that includes the internal criteria of control described in this section, has been proposed by myself and is based on the inclusion of molecules with electrons within the classical atomistic/coarse-grained Adaptive Resolution scheme (AdResS) in its Grand Canonical version (GC-AdResS) \cite{cpcluigi}. The essence of the idea is reported in the next section.
 \subsection{el-QM-AdResS: A Grand Canonical electronic system embedded in a classical reservoir}
\begin{figure}
   \centering
   \includegraphics[width=0.75\textwidth]{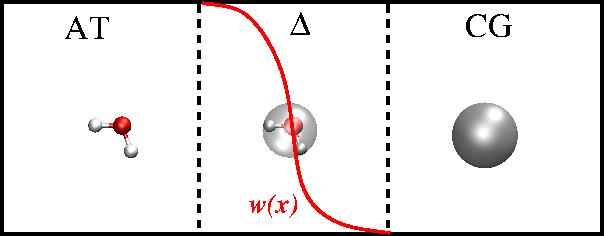}
   \caption{Schematic representation of the AdResS simulation set up. AT indicates the atomistic region; $\Delta$ is the region where molecules change resolution according  to $w(x)$, and CG is the coarse-graining region. {Adapted with permission from Figure 1 of Ref.\cite{cpcluigi}. License number 4355330763152, Copyright 2018, Elsevier}}
 \label{cartoon-adress}
 \end{figure}

\begin{figure}[h!]
  \centering
  \includegraphics[width=0.80\textwidth]{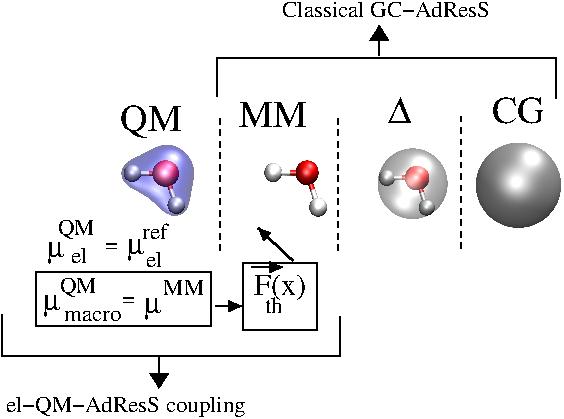}
  \caption{Pictorial representation of the el-QM-AdResS. ${\mu}_{macro}^{QM}$ is the macroscopic chemical potential assured by the thermodynamic force in the MM and the $\Delta$ region. In essence, as in the $\Delta$ region of a purely classical GC-AdResS, an iterative procedure is applied so that the average number density of molecules in the MM region (and of course in the $\Delta$ and CG regions of el-QM-AdResS) is equal to the molecule number density of reference ({at which the QM region automatically is}). ${\mu}_{el}^{QM}=\mu_{el}^{ref}$ is the electronic chemical potential of the QM region (which should be the same as that of a corresponding full quantum calculation of the bulk/environment).}
 \label{el-qm-ad-scheme}
\end{figure}

 The Adaptive Resolution Simulation technique (AdResS) \cite{jcp,annurev} for classical systems allows molecules to transform their resolution according to the region in space where they are instantaneously located. The root model of the computational algorithm consists of a  space dependent interpolation for the force between two molecules, $\alpha,\beta$ (see Fig.\ref{cartoon-adress}):
\begin{equation}
F_{\alpha \beta} = w(X_{\alpha})w(X_{\alpha})F_{\alpha\beta}^{AT} + [1 - w(X_{\alpha})w(X_{\alpha})]F_{\alpha\beta}^{CG}
\end{equation}  
where $F_{\alpha\beta}^{AT}$ is the atomistic force and $F_{\alpha\beta}^{CG}$ is the coarse-grained force.  
$w(x)$ ($x$ is the coordinate of the center of mass of a molecule) smoothly goes from 0 to 1 in a transition region $\Delta$. The system is embedded in a thermostat that assures a target thermodynamic equilibrium. The conceptual solidity of the original AdResS method is then enhanced by the addition of a force, ${\bf F}_{th}(x)$ (thermodynamic force), which acts on the center of mass of the molecule in $\Delta$. Such a force assures that, in situations of equilibrium, the effective chemical potential of the whole system corresponds to that of the (target) atomistic resolution \cite{jcpsimon,prx,njp}. ${\bf F}_{th}(x)$, has been derived within a rigorous statistical mechanics model for molecular systems with open boundaries \cite{prl12,jctchan,prx,mujcp,njp,pre16,physrep} (thus AdResS became Grand Canonical (GC-) AdResS) and it was found out that a numerically convenient way to calculate ${\bf F}_{th}(x)$ during a simulation (equilibration run), consists of expressing it as the gradient of the number density of the molecules in an iterative form \cite{prl12}: $F_{k+1}^{th}(x)=F_{k}^{th}(x) - \frac{M_{\alpha}}{[\rho_{ref}]^2\kappa}\nabla\rho_{k}(x)$,
$M_{\alpha}$ is the mass of the molecule, $\kappa$ a (conveniently) tunable constant, $\rho_{k}(x)$ is the molecular density  as a function of the position in $\Delta$ at the $k$-th iteration and $\rho_{ref}$ is the density of reference, decided {\it a priori} according to the thermodynamic state point at which we wish to do the simulation.
The convergence criterion depends on the accuracy required for the simulation
but, based upon experience, $|\rho_{final}-\rho_{ref}|$ should always be below $10\%$ in $\Delta$. The accuracy of this method has been proven over the last ten years over a large range of systems and problems, see e.g. \cite{physrep} for an overview, and Refs.\cite{jcpil,advts} for the most recent applications. In this context, it is important to notice that the coupling scheme allows one to interface any classical molecular representations (e.g. two different atomistic models, atomistic and path integral representation of molecules \cite{lujcppi,cpcanim}, to mention a few). There also exists a version based on a global Hamiltonian \cite{raff1,raff2}, which is, in essence, technically equivalent to the method outlined above (although conceptually confusing, see discussion in \cite{physrep}). Given the technical equivalence, it is not a surprise that calculations repeated with the Hamiltonian scheme give the same results obtained years before with the (GC-)AdResS method (compare \cite{lujcppi,cpcanim}, with \cite{raffpi1,raffpi2}; \cite{matdna} with \cite{raffdna}; and \cite{mujcp} with \cite{raffmu}).\\ 
In order to include electrons, the theoretical framework of el-QM-AdResS, proposed in Reference \cite{cpcluigi}, is based on two simple concepts (see also pictorial representation in Figure \ref{el-qm-ad-scheme}):
\begin{itemize}
\item Use the classical GC-AdResS scheme as a computational interface for the quantum region (i.e. equivalent to the MM region of a QM/MM scheme).\\
  GC-AdResS  will take care of providing and removing the (classical) nuclei (i.e. the skeleton of the molecule) in the QM region according to the macroscopic (thermodynamic) chemical potential.
\item The electrons of the QM region are treated as a system in contact with an ideal (Grand Canonical) reservoir of electrons whose exchange of particles is regulated by a properly chosen (see explanation later) electronic chemical potential.
\end{itemize}
The scheme follows three automatic criteria of control for assuring physical consistency of the calculation:
\begin{itemize}
\item The coupling between the QM region and the classical reservoir must be such that: $\langle H_{QM-Ad}\rangle << \langle H_{QM}\rangle$ at any time, with $ H_{QM-Ad}$ the coupling Hamiltonian of the QM region with the GC-AdResS interface (as an example of application of this criterion in GC-AdResS, see \cite{njp,lujcppi}).
\item The macroscopic chemical potential must be such that $\mu_{macro}^{QM}=\mu_{macro}^{GC-AdResS}$, i.e. the whole system is at the same macroscopic chemical potential at any time.
\item The electronic chemical potential in the QM region must satisfy the condition: $\mu_{el}^{QM}=\mu_{el}^{ref}$, where $\mu_{el}^{ref}$ is the electronic chemical potential the QM region would have if the whole system was treated at quantum resolution. For example, for a molecule solvated in water, the QM region would be composed of the solute and the water molecules of the first solvation shells embedded in bulk water, thus the electronic chemical potential of reference would be that of bulk water.
\end{itemize}
The first condition is a simple check that assures the choice of a physically consistent size of the QM region. The second condition is fulfilled by the thermodynamic force, calculated with the same formula used in standard GC-AdResS for the region $\Delta$, but now calculated for (and applied in) the MM region at the interface with the QM region. Finally, the third condition can be implemented by performing electronic structure calculations for the QM region at constant $\mu_{el}$, with the number of electrons being the variable in the (Grand Canonical) energy functional minimization (in the same spirit of the Grand Canonical DFT calculations of the previous sections). The electronic chemical potential can be determined from separate calculations of prototype systems that represent a reasonable bulk/environment (see example of liquid water before).  The main advantage of the el-QM-AdResS scheme consists in the assurance regarding the statistical mechanical consistency of the results; however, {it carries with it open questions regarding the integration of its conceptual aspects and its technical implementation.} In fact, from the technical point of view, one major challenge is the abrupt interface between the QM and the GC-AdResS region. It is possible that the thermodynamic force, acting in the MM region directly adjacent to the QM region which, in essence, equalizes the molecular number density in an iterative process (equilibration run), may not converge as expected. It is encouraging though that recent tests have proved that this may not be a problem; in fact, an abrupt interface can be turned into an advantage \cite{prlyn}. Another challenge is the fact that $N$, the number of electrons in the QM region, may not be integer, which implies that some fractional charge is delocalized in the classical region and it must be taken into account. In the meantime, some technical solutions have been proposed \cite{cpcluigi} and probably the optimal development of this idea is to combine it with the already well developed technical A-QM/MM schemes. A brief discussion about this possibility is considered in the next section.
 \subsection{el-QM-AdResS + A-QM/MM: A path to a technically powerful and conceptually self-validating innovative scheme}
 el-QM-AdResS is, as a matter of fact, technically similar to A-QM/MM, and some work in merging AdResS with A-QM/MM has been already made in the group of Rosa Bulo \cite{bulo-adr}. A proposed way to proceed further would be to extend the criteria of physical validity used in el-QM-AdResS to A-QM/MM  and, from the other side, use a buffer region and the corresponding partitioning scheme of A-QM/MM in el-QM-AdResS. The resulting method would be a new A-QM/MM where each extended QM region, corresponding to each partition, is treated within the (same) $\mu_{el}$ constant scheme (and implicitly, through the thermodynamic force, at the same $\mu_{macro}$). As a consequence, the average potential (or forces) calculated over all partitioning schemes in the buffer region would correspond to an average over different, but equivalent (i.e. same $\mu_{el}$ and same $\mu_{macro}$), Grand Canonical systems, and thus, it would have a solid statistical mechanics justification. At the same time the technical problem of the abrupt interface in el-QM-AdResS could be solved by the smoothing technique of the buffer region, where the thermodynamic force can also be applied and smoothly calculated. Moreover, GC-AdResS allows a further coupling of the atomistic resolution to coarse-grained models and beyond (continuum models) \cite{matej1,matej2}, thus, the fully-realized A-QM/MM approach would represent a truly multiscale method.\\
 Most of the QM calculations are done within the DFT scheme, however, one can, in principle, go beyond DFT and aim for a higher level of accuracy. For example, Hofer and H\"{u}nenberger have recently explored the idea of going beyond DFT and using the resolution-of-identity second-order M{\o}ller-Plesset perturbation (RIMP2), which is a sort of first level beyond DFT for describing electron correlations \cite{hunen}.  In this perspective, it is important to also consider other electronic structure methods that can accurately describe electron correlations but, given the context of this work, also assure exchange of matter, and thus calculations involving a varying number of electrons. In the previous chapters, quantum chemistry approaches in Fock space have already been discussed, however, one may also extend the discussion to other methods such as Quantum Monte Carlo (QMC). As previously anticipated, QMC techniques in the Grand Canonical ensemble are developed and used mostly for technical reasons (alleviating size effects), but in any case, in this context, they may be included in a scheme such as el-QM-AdResS or A-QM/MM. A discussion of QMC methods where the number of electrons can vary is reported in the next section.
\section{Quantum Monte Carlo {and a} varying number of electrons}
Quantum Monte Carlo methods can be classified among the most accurate techniques for describing electron correlations. There are multiple QMC approaches, each with its own positive and negative aspects; however, in general they are all closely related. {Here, I will not provide an overview of specific QMC techniques (for an updated discussion see Reference \cite{qmc-generic} and references therein) but I will instead focus on some generic features that can be applied to different QMC approaches and involve the possibility of carrying out QMC calculations in a Grand Canonical ensemble}. The work of reference which I will follow is the paper by Lin, Zong and Ceperley who introduced the so-called ``twist average Grand Canonical ensemble'' (TA-GCE) \cite{qmc-size1}.\\
TA-GCE has been developed in order to reproduce results of the thermodynamic limit with calculations involving a (relatively small) finite number of electrons, thus circumventing finite size effects. The starting point of this idea is the observation that in QMC the phase factor of the many-electron wavefunction  is usually assumed to return to the same value if {electrons} cross the periodic boundary and return to the same position. The technical consequence of this assumption is that, for delocalized electrons, the convergence of properties to their thermodynamic limit is very slow.  In Reference \cite{qmc-size1}, Lin, Zong and Ceperley proposed an alternative that allows {electrons} to {take up a phase angle} when they cross the periodic boundary. This is the basic concept of twist averaged boundary conditions (TABC); the pre-existing literature background they refer to is a work by Gros \cite{qmc-size-a} who has shown that TABC, applied to the Hubbard model, gives exact results in the Grand Canonical ensemble for non-interacting particles. Thus, for the focus of this paper, this is a procedure of interest for the treatment of a varying number of electrons. In the following, I will report the basic ingredients of the idea of Lin, Zong and Ceperley. In non-interacting homogeneous systems with periodic boundary conditions, plane waves describe single particle states. For simplicity of, exposition the spinless case is considered, so that the single state of the system in a, e.g., cubic box of linear size $L$, can be written as: $\phi({\bf r}, {\bf k})\propto e^{i{\bf k}{\bf r}}$. In order to satisfy the TABC one must have: ${\bf k}_{\bf n}=(2\pi+\theta)/L$, with ${\bf n}$ an integer vector. In the Canonical ensemble, the $N$ lowest energy states form the ground state and a Slater determinant of such states gives the ground state wavefunction of the system. In the Grand Canonical ensemble one has that: $E_{\alpha,N}=E_{\alpha,N}(\theta)$, where $(\alpha,N)$ labels the quantum states. The probability of a given state is: $P(\alpha,N)\propto exp\{-\beta[E_{\alpha,N}(\theta)-N\mu]\}$, with $\mu$ the chemical potential of the system. In the ground state, that is for $\beta\to \infty$, the wavefunction must not optimize the energy $E_{\alpha,N}$, as in the Canonical ensemble, but the quantity: $E_{\alpha,N}(\theta)-\mu N$. Next, they show that the occupation number is exactly what one gets if the calculation were to be done in the thermodynamic limit, and thus TABC (now renamed TA-GCE) carries no size effects. The procedure implies that the number of {electrons} varies, since, for a given $\theta$ and Fermi wave vector, the number of occupied states will vary. The fluctuations in the particle number is then derived and it is shown to be $\propto N^{1/4}$ in the limit for $N\to \infty$. The treatment discussed so far concerns non-interacting systems; the extension to interacting systems is based on the argument that the theory of Fermi Liquids states that low lying excited states of a non-interacting system are in a one-to-one correspondence with non-interacting states, thus TA-GCE is likely to reduce finite size effects for interacting systems, as well. There are technical difficulties, though, that must be reported. One is that the wavefunction must be optimized at each value of $\theta$, which implies additional computational resources. Another problem is the treatment of net charges that are not balanced, but one can show that the average
charge of the supercell (averaged over the BC) is exactly given by the density; so, in many cases, this is sufficient to get rid of the problem of
a charged supercell. In general, reasonable solutions to such difficulties have been found, and the method has been applied to calculate several properties of interacting electrons \cite{qmc-size2,markus1,markus2,markus3}. In the schematic picture of {\it ``system/environment''},  with which I have analyzed methods and applications so far, the TA-GCE approach can be visualized as the construction of a generic ``infinite'' environment (reservoir of the Grand Canonical ensemble) via the introduction of the phase factor of the wavefunction that breaks the symmetry of the periodicity. This characteristic suggests that in principle, TA-GCE, in the not too distant future, may find use beyond its current technical utility of circumventing finite size effects. In this perspective, TA-GCE could be employed, for example, to describe the (small) QM part of A-QM/MM or el-QM-AdResS treated in the previous section, or for the QM description of the electrode in electrochemistry. Besides TA-GCE, other routes have been explored within the QMC scheme. An interesting recent example concerns the possibility of considering systems with a variable number of electrons for studying photoemission and inverse photoemission phenomena \cite{photo}. The method used is the Full Configuration Interaction QMC method (FCIQMC) developed in the group of Ali Alavi \cite{fciqmc1,fciqmc2,fciqmc3}. In essence, the method uses a Monte Carlo sampling of Slater determinants and calculates the real-time propagation of the wave function of the system as an
electron is added or removed. Differently from TA-GCE, electrons are not introduced in a Grand Canonical fashion, but the method is specifically designed to compute electronic spectra upon sudden addition or
removal of an electron (as in photoemission). In a qualitative explanation, this is equivalent to saying that the removed electron has been put in a stationary, non-interacting state somewhere at infinity. It must be added that nowadays it is even possible to run small QMC calculations on desktop {computers; however, the computational costs of any QMC approach for large-scale systems routinely treated by, e.g., DFT, requires high-performance resources that may be beyond the means of research groups of a moderate size. Nevertheless, or, actually, because of such limitations, there is the need to significantly intensify the amount of theoretical work being done in this field}.
\section{Conclusions and Perspectives}
The overall analysis of the various situations discussed in this paper leads me to the conclusion that {what lies at the core of the problem that is simulating open, electronic systems,} is not the lack of accurate theoretical tools, but rather the difficulty of merging the corresponding simulation techniques, specific to multiple aspects of the problem, in a physically consistent way. A proper, non-empirical merging is a necessary step for building simulation tools with predictive power, without the mandatory need for a case-by-case external validation. In the last decade we have witnessed the exponential increase of multiscale simulation techniques, mostly based on pragmatic empiricism for solving specific problems in the short term. The challenge of the next decade, in my view, should be the construction of multiscale techniques where the scale-coupling interfaces are based only on solid physical principles with corresponding well defined formulas for possible errors induced by approximations. Such a project requires methodological and theoretical work with, as much as possible, mathematical rigor, thus strengthening even further the concept that methodology (across fields and disciplines) is a self-standing, relevant field. On a broader scale, the project would require a combined effort of applied mathematicians, theoretical physicists and simulators (including physicists, chemists and materials scientists) with constructive exchanges is every direction.
In this perspective, the aim of this paper is to offer to the molecular simulation and related communities a bird's-eye view (of a theoretical physicist) on recent progress in the computational treatment of many-electron systems embedded in a fully-open larger environment. This paper shall not be intended as a comprehensive review but rather as a stimulating discussion for future developments, taking, as possible starting points, the methods, ideas and applications discussed in the various sections. These examples discussed have not been chosen because they are necessarily more relevant than others, but because together they sample the field in a rather uniform way. Their overall analysis leads to the conclusion that, as a QM technique, DFT clearly plays a key role, however quantum chemical methods (beyond Hartree-Fock) and QMC methods have been developed, at least at a conceptual level, up to the point of allowing calculations at a varying number of electrons that mimic the exchange of matter with the environment. Yet, their computational price is still prohibitive for treating systems such as molecules in solution or the metal surfaces associated with electrochemistry. The hope is that with the rapid evolution of the technology related to computational capabilities, the application of such methods to the above-mentioned systems may become possible. An intermediate step, that I can foresee, may be represented by the development of methods similar to the QM/MM approach with open boundaries, but with different levels of QM resolution (for a similar idea within the density matrix embedding approach see \cite{knizia}). {An example could be Grand Canonical QMC embedded in a quantum reservoir treated at the (Grand Canonical) DFT level, itself embedded in a reservoir of classical molecules. Analogous to the coupling criteria of el-QM-AdResS, one would require that the electronic and macroscopic chemical potentials are equal at the various interfaces.} In summary, the future challenge lies in the construction of a solid theory of boundary conditions, at the interface of varying resolutions, in order to achieve a global physical consistency. A selection of starting points has been reported here and these are now ready to be developed even further.

\section*{Acknowledgments}{This research has been funded by Deutsche Forschungsgemeinschaft (DFG) through the grant CRC 1114: ``Scaling Cascades in Complex Systems'', project C01. I would like to thank Paul Ayers, Ferdinand Evers and Christian Krekeler for useful suggestions that have clarified relevant concepts discussed in this work. I am grateful to Ali Alavi, David Ceperley, Markus Holzmann and Ravishankar Sundararaman, for a critical reading of the sections of the manuscript related to their field of expertise. {I wish to thank John Whittaker for a critical reading of the manuscript and for his precious suggestions.}}


\begin{thebibliography}{24}
\expandafter\ifx\csname natexlab\endcsname\relax\def\natexlab#1{#1}\fi
\expandafter\ifx\csname url\endcsname\relax
  \def\url#1{\texttt{#1}}\fi
  \expandafter\ifx\csname urlprefix\endcsname\relax\def\urlprefix{URL }\fi
\bibitem{physrep}
  L. Delle Site and M.Praprotnik, Phys. Rep. (2017) {\bf 693}, 1-56
\bibitem{kais}
  Z.Huang and S.Kais, Chem.Phys.Lett. (2005), {\bf 413}, 1
\bibitem{markusreiher}
 K. Boguslawski, P. Tecmer, \"{O}. Legeza and M.Reiher, J.Phys.Chem.Lett. (2012), {\bf 3}, 3129
\bibitem{tecmer}
 K.Boguslawski and P.Tecmer, Int.J.Quant.Chem. (2014), {\bf 115}, 1289 
\bibitem{ijqc}
  L.Delle Site, Int.J.Quant.Chem. (2015) {\bf 115}, 1396
\bibitem{dft}
  P.Hohenberg and W.Kohn, Phys.Rev. (1964) {\bf 136}, A1133; W.Kohn and L.Sham, Phys.Rev. (1965) {\bf 140}, A1441
\bibitem{bookdft}
W. Parr and R.G. Yang,
{\it Density-Functional Theory of Atoms and Molecules}
Oxford University Press (1989)
\bibitem{qmmmad2}
M. Zheng and M. P. Waller, WIREs Comp.Mol.Sci. (2016), {\bf 6}, 369
\bibitem{qmmmad}
  A.W.Duster, C.-H. Wang, C.M.Garza, D.E.Miller and H.Lin, WIREs Comp.Mol.Sci. (2017) {\bf 7}, e1310  
\bibitem{cpcluigi}
  L.Delle Site, Comp.Phys.Comm. (2018) {\bf 222}, 94
\bibitem{lozovoi}
  A.Y Lozovoi, A.Alavi, J.Kohanoff and R.M. Lynden-Bell, J.Chem.Phys. (2001) {\bf 115}, 1661
\bibitem{sprik}
  I.Tavernelli, R. Vuilleumier and M.Sprik, Phys.Rev.Lett. (2002) {\bf 88}, 213002
\bibitem{anatole}
  O. von Lilienfeld and M.Tuckerman, J.Chem.Phys. (2006) {\bf 125}, 154104
\bibitem{auer}
  W.B.Schneider and A.A.Auer, Beil.Jour.Nanotech. (2014) {\bf 5}, 668
\bibitem{arias}
 R.Sundararaman, W.A.Goddard III and T.A.Arias, J.Chem.Phys. (2017) {\bf 146}, 114104
\bibitem{ayers1}
  R.A.Miranda-Quintana, P.K.Chattaraj and P.W.Ayers, J.Chem.Phys. (2017) {\bf 147}, 124103
\bibitem{bochum}
W.Kutzelnigg, (1989){\it Quantum Chemistry in Fock Space}, in: Mukherjee D. (eds) {\it Aspects of Many-Body Effects in Molecules and Extended Systems.} Lecture Notes in Chemistry, vol 50. Springer, Berlin, Heidelberg  
\bibitem{qmc-generic}
  M.A.Morales, R.Clay, C.Pierleoni and D.M.Ceperley, Entropy (2014) {\bf 16}, 287
\bibitem{qmc-size-a}
  C.Gros, Phys.Rev.B (1996) {\bf 53}, 6865
\bibitem{qmc-size-b}
 A.N.Rubtsov and A.I.Lichtenstein, JETP Lett. (2004) {\bf 80}, 61
\bibitem{qmc-size1}
  C.Lin, F.H.Zong and D.M.Ceperley, Phys.Rev.B (2001) {\bf 64}, 016702
\bibitem{qmc-size2}
  S. Chiesa, D.M. Ceperley, R.M. Martin, and M. Holzmann, Phys.Rev.Lett (2009) {\bf 97}
\bibitem{nanoel1}
  C.-Y. Lai and C.-C. Chien, Sci.Rep. (2016), {\bf 6}, 37256
\bibitem{nanoelnic}
  P.B.Coto, C.Hofmeister, V.Prucker, D.Weckbecker and M.Thoss, {\it Simulation of Electron Transfer and Electron Transport in Molecular Systems at Surfaces}, in NIC Symposium 2016, K.Binder, M.M\"{u}ller, M.Kremer, A.Schnurpfeil (Eds.), Schriften des Forschungszentrums J\"{u}lich, NIC series Vo.48, pp.133
\bibitem{nanoelrefpap}
 M.Thoss and F.Evers, J.Chem.Phys. (2018), {\bf 148}, 030901
\bibitem{nanoel-advts}
  V.Diez-Cabanes, S.Rodriguez Gonzalez, S.Osella, D.Cornil, C.Van Dyck and J.Cornil, Adv.Theory Simul.(2018), DOI: 10.1002/adts.201700020
\bibitem{46ofarias}
 E.Herrero, L.J.Buller and H.D.Abruna, Chem.Rev. (2001), {\bf 101}, 1897
\bibitem{spcnano}
F.Evers and L.Venkataraman, J.Chem.Phys. (2017), {\bf 146}, 092101
\bibitem{landauer}
R. Landauer, IBM J. Res. Dev. (1957), {\bf 1}, 223
\bibitem{ratner}
  Y.Xue and M.A.Ratner, Phys.Rev.B (2003), {\bf 68}, 115406
\bibitem{koentopp}
F.Evers, F.Weigend, and M.Koentopp, Phys.Rev.B, (2004), {\bf 69}, 235411
\bibitem{quasi-gc}
  A.Arnold, F.Weigend and F.Evers, J.Chem.Phys. (2007), {\bf 126}, 174101
\bibitem{emsew}
  G.G.Emch and G.L.Sewell, J.Math.Phys. (1968), {\bf 9}, 946
\bibitem{zwanzig}
  R.Zwanzig, J.Chem.Phys. (1960), {\bf 33}, 1338; R.Zwanzig, Physica (1964), {\bf 30}, 1109
\bibitem{memory1}
  G.Cohen, E.Wilner and E.Rabani, New J.Phys. (2013), {\bf 15}, 073018
\bibitem{memory2}
E.Wilner, H.Wang, M.Thoss and E.Rabani, Phys.Rev.B, (2014) {\bf 89}, 205129  
\bibitem{linda}
 G.Lindblad, Comm.Math.Phys. {\bf 48}, 119 (1976)
\bibitem{kosslinda}
  A.Kossakowski,  Rep. Math. Phys. {\bf 3}, 247 (1972)
\bibitem{kosov}
 D.S. Kosov, (2009), J.Chem.Phys.{\bf 131}, 171102
\bibitem{neuhauser}
I.V.Ovchinnikov and D.Neuhauser, J.Chem.Phys. (2005), {\bf 122}, 024707
\bibitem{tamar}
  T.Zelovich, T.Hansen, Z.-F.Liu, J.B.Neaton, L.Kronik and O.Hod, J.Chem.Phys. (2017), {\bf 146}, 092331
\bibitem{thosswf}
  H.Wang and M.Thoss, J.Chem.Phys. (2009), {\bf 131}, 024114
\bibitem{lubic}
  C.Lubic, Math.Comp. (2005), {\bf 74}, 765
\bibitem{vanleuven}
  J.Broekhove, L.Lathouwers, E.Kesteloot and P.Van Leuven, Chem.Phys.Lett. (1988), {\bf 149}, 547
\bibitem{thosswf2}
  H.Wang and M.Thoss, J.Chem.Phys. (2016), {\bf 145}, 164105
\bibitem{kvaal}
  S.Kvaal, Phys.Rev.A (2011), {\bf 84}, 022512
\bibitem{gor}
  V.Gorini, A.Kossakowski and E.Sudarshan, J.Math.Phys. (1976), {\bf 17}, 821
\bibitem{cat1}
  P. De Luna, R.Quintero-Bermudez, C.-T.Dinh, M.B. Ross, O.S. Bushuyev, P.Todorovic, T.Regier, S.O. Kelley, P.Yang and E.H. Sargent, Nat.Cat. (2018), {\bf 1}, 103
\bibitem{ali}
  A.Alavi, J.Kohanoff, M.Parrinello and D.Frenkel, Phys.Rev.Lett. (1994), {\bf 73}, 2599
\bibitem{thomas}
  D.Richters and T.D. K\"{u}hne, J.Chem.Phys. (2014), {\bf 140}, 134109
\bibitem{mermin}
  N.D.Mermin, Phys.Rev. (1965), {\bf 137}, A1133
\bibitem{parr1}
F. R. Krajewski and M. Parrinello, Phys. Rev. B (2005), {\bf 71}, 233105
\bibitem{parr2}
  F. R. Krajewski and M. Parrinello, Phys. Rev. B (2006), {\bf 73}, 041105
\bibitem{perdew}
  J.P.Perdew, R.G.Parr, M.Levy, J.L.Balduz, Phys.Rev.Lett. (1982), {\bf 49}, 1691
\bibitem{vuilleumier}
  R.Vuilleumier, M.Sprik and A.Alavi, J.Mol.Struct: THEOCHEM  (2000), {\bf 506}, 343
\bibitem{weitao}
 C.Li, J.Lu and W.Yang, J.Chem.Phys. (2017), {\bf 146}, 214109
\bibitem{cont1}
R.Sundararaman and T.A.Arias, Comp.Phys.Comm. (2014), {\bf 185}, 818
\bibitem{cont2}
 R.Sundararaman, K.Letchworth-Weaver and T.A.Arias, J.Chem.Phys. (2014), {\bf 140}, 144505
\bibitem{jdft1}
 R.Sundararaman, K.Schwarz, K.Letchworth-Weaver and T.A.Arias, J.Chem.Phys. (2015), {\bf 142}, 054102
\bibitem{jdft2}
  R.Sundararaman and W.A.Goddard III, J.Chem.Phys. (2015), {\bf 142}, 064107
\bibitem{bulo-matter}
  J.M. Boereboom, P.Fleurat-Lessard and R.E. Bulo, J.Chem.Th.Comp. (2018), DOI: 10.1021/acs.jctc.7b01206
\bibitem{warschkarplus}
  A. Warshel, and M.Karplus,  J. Am. Chem. Soc. (1972), {\bf 94}, 5612
\bibitem{birge}
 R.R. Birge, M.J. Sullivan, B.E. Kohler, J. Am. Chem. Soc. (1976), {\bf 98}, 358
\bibitem{warshlev}
  A.Warshel, and M.Levitt, J. Mol. Biol. (1976), {\bf 103}, 227
\bibitem{roth}
  E.Brunk, and U.Rothlisberger, Chem.Rev. (2015), {\bf 115}, 6217
\bibitem{thiel}
  H.Senn, and W.Thiel, Angew. Chem. Int. Ed. (2009), {\bf 48}, 1198
\bibitem{jcp}
M.Praprotnik, L.Delle Site, and K.Kremer, J.Chem.Phys.  (2005), {\bf 123}, 224106 
\bibitem{annurev}
M.Praprotnik, L.Delle Site, and K.Kremer, Annu.Rev.Phys.Chem. (2008), {\bf 59}, 545 
\bibitem{prx}
H.Wang, C.Hartmann, C. Sch\"{u}tte and L.Delle Site, Phys.Rev.X, (2013) {\bf 3}, 011018
\bibitem{njp}
 Agarwal and J. Zhu and C. Hartmann and H. Wang and L.Delle Site, New. J. Phys. (2015), {\bf 17}, 083042
\bibitem{heydenqmmm}
  A.Heyden, H.Lin, and D.G.Truhlar, J.Phys.Chem. B, (2007), {\bf 111}, 2231
\bibitem{bulos1}
  R.Bulo, B.Ensing, J.Sikkema, and L.Visscher, J. Chem. Theory. Comput. (2009), {\bf 5}, 2212
\bibitem{nielsen}
  S.Nielsen, R.Bulo, P.Moore, and B.Ensing, Phys. Chem. Chem. Phys. (2010), {\bf 12}, 12401
\bibitem{csanys}
N.Bernstein, C.Varnai, I. Solt, S.Winfield, M.Payne, I. Simon, M.Fuxreiter, G. Csany, Phys. Chem. Chem. Phys. (2012), {\bf 14}, 646
\bibitem{watanabe}
  H. Watanabe, T.Kubar, and M.Elstner, J. Chem. Theory Comput. (2014), {\bf 10}, 4242
\bibitem{lins1}
  S.Pezeshki, and H.Lin, J. Chem. Theory Comput. (2014), {\bf 10}, 4765
\bibitem{lins2}
  S.Pezeshki, C.Davis, A.Heyden, and H.Lin, J.Chem. Theory Comput. (2014), {\bf 10}, 476
  {  \bibitem{nearkohn}
 W.Kohn, Phys.Rev.Lett. (1996), {\bf 76}, 3168 
\bibitem{nearkohn2}
  E.Prodan, and W.Kohn, PNAS, (2005), {\bf 102}, 11635
\bibitem{ayers-gerl}
S.Fias, F.Heidar-Zadeh, P.Geerlings, and P.W. Ayers, PNAS, (2017), {\bf 114}, 11633}
\bibitem{quintana-ayers}
  R.A.Miranda-Quintana, and P.W.Ayers, J.Chem.Phys. (2016) {\bf 144}, 244112
\bibitem{jcpsimon}
  S.Poblete, M.Praprotnik, K.Kremer and L.Delle Site, J.Chem.Phys. (2010), {\bf 132}, 114101
\bibitem{prl12}
  S.Fritsch, S.Poblete, C.Junghans, G.Ciccotti, L.Delle Site and K.Kremer, Phys.Rev.Lett. (2012), {\bf 108}, 170602
\bibitem{jctchan}
  H.Wang, C.Sch\"{u}tte and L.Delle Site, J.Chem.Th.Comp. (2012), {\bf 8}, 2878
\bibitem{mujcp}
  A.Agarwal, H.Wang, C.Sch\"{u}tte and L.Delle Site, J.Chem.Phys. (2014), {\bf 141}, 034102
\bibitem{pre16}
  L.Delle Site, Phys.Rev.E {\bf 93}, (2016), 022130
\bibitem{jcpil}
  B.S.Jabes, C.Krekeler, R.Klein, and L.Delle Site, J.Chem.Phys (2018), {\bf 148}, 193804
\bibitem{advts}
  B.S.Jabes, R.Klein, and L.Delle Site, Adv.Th.Sim. (2018), in press
\bibitem{lujcppi}
  A.Agarwal, and L.Delle Site, J.Chem.Phys. (2015), {\bf 143}, 094102
\bibitem{cpcanim}
  A.Agarwal and L.Delle Site, Comp.Phys.Comm. (2016), {\bf 206}, 26
\bibitem{raff1}
  R.Potestio, S.Fritsch, P.Espanol, R.Delgado-Buscalioni, K.Kremer, R.Everaers and D.Donadio, Phys.Rev.Lett. (2013), {\bf 110}, 108301
\bibitem{raff2}
  R.Potestio, P.Espanol, R.Delgado-Buscalioni, K.Kremer, R.Everaers and D.Donadio, Phys.Rev.Lett. (2013), {\bf 111}, 060601
\bibitem{raffpi1}
  K.Kreis, M.E. Tuckerman, D.Donadio, K.Kremer and R.Potestio, J. Chem. Theory Comput. (2016), {\bf 12}, 3030
\bibitem{raffpi2}
  K.Kreis, K.Kremer, R.Potestio, and M.E. Tuckerman, J.Chem.Phys. (2017), {\bf 147}, 244104
\bibitem{matdna}
  J.Zavadlav, R.Podgornik and M.Praprotnik, J. Chem. Theory Comput. (2015), {\bf 11}, 5035
\bibitem{raffdna}
T.Tarenzi, V.Calandrini, R.Potestio, A.Giorgetti, and P.Carloni, J.Chem.Th.Comp. (2017), {\bf 13}, 5647 
\bibitem{raffmu}
  M.Heidari, K.Kremer, R.Cortes-Huerto, and R. Potestio, (2018), arXiv preprint arXiv:1802.08045
\bibitem{prlyn}
  C.Krekeler, A.Agarwal, C.Junghans, M.Praprotnik, and L.Delle Site, J.Chem.Phys. (2018) submitted
\bibitem{bulo-adr}
  J.M.Boereboom, R.Potestio, D.Donadio, and R.Bulo, J.Chem.Th.Comp. (2016), {\bf 12}, 3441
\bibitem{matej1}
R.Delgado-Buscalioni, K.Kremer, and M. Praprotnik, J.Chem.Phys. (2008), {\bf 128}, 114110
\bibitem{matej2}
  R.Delgado-Buscalioni, K.Kremer, and M. Praprotnik, J.Chem.Phys. (2009), {\bf 131}, 244107
\bibitem{hunen}
  T.S.Hofer, and P.H. H\"{u}nenberger, J.Chem.Phys. (2018), {\bf 148}, 222814
\bibitem{markus1}
M. Holzmann, B. Bernu, V. Olevano, R.M. Martin, and D.M. Ceperley, Phys. Rev. B (2009) {\bf 79}, 041308(R)  
\bibitem{markus2}
M. Holzmann, B. Bernu, C. Pierleoni, J. McMinis, D. M. Ceperley, V. Olevano, and L. Delle Site, Phys. Rev. Lett. (2011), {\bf 107}, 110402  
\bibitem{markus3}
  M. Holzmann, R.C. Clay III, M. A. Morales, N.M. Tubman, D. M. Ceperley, and C. Pierleoni, Phys. Rev. B (2016), {\bf 94}, 035126
\bibitem{fciqmc1}
  G.H.Booth, A.J.W.Thom, and A.Alavi, J.Chem.Phys. (2009), {\bf 131}, 054106
 \bibitem{fciqmc2}
D.Cleland, G.H.Booth, and A.Alavi, J.Chem.Phys. (2010), {\bf 132}, 041103 
\bibitem{fciqmc3}
  G.H.Booth, and A.Alavi, J.Chem.Phys. (2010), {\bf 132}, 174104
\bibitem{photo}
  K.Guther, W.Dobrautz, O.Gunnarson, and A.Alavi, (2017) arXiv:1709.00218v1
\bibitem{knizia}
G.Knizia, and G.K.-L.Chan, J.Chem.Th.Comp. (2013), {\bf 9}, 1428
\end{thebibliography}
\end{document}